\documentclass[slac_one]{revtex4}
\usepackage{graphicx}
\usepackage{fancyhdr}
\newcommand{\non}{\nonumber}
\def\gtsim{\mathrel{\hbox{\raise0.2ex
\hbox{$>$}\kern-0.75em\raise-0.9ex\hbox{$\sim$}}}}
\def\ltsim{\mathrel{\hbox{\raise0.2ex
\hbox{$<$}\kern-0.75em\raise-0.9ex\hbox{$\sim$}}}}
\begin{document}
\rightline{OU-HET 537}
\rightline{KEK-TH-1028}
%
\title{{\small{2005 International Linear Collider Workshop - Stanford,
U.S.A.}}\\ 
\vspace{12pt}
Electroweak baryogenesis and the triple Higgs boson coupling} 
\author{Shinya~Kanemura}
\affiliation{Osaka University, Toyonaka, Osaka 560-0043, Japan}
\author{Yasuhiro~Okada}
\affiliation{KEK, Tsukuba, Ibaraki 305-0801, Japan}
\author{Eibun~Senaha}
\affiliation{KEK, Tsukuba, Ibaraki 305-0801, Japan}
\begin{abstract}
We study collider signatures of electroweak baryogenesis in the two Higgs doublet model
and the minimal supersymmetric standard model.
It is found that the trilinear coupling of the 
lightest Higgs boson receives large quantum corrections if the electroweak phase 
transition is strongly first order for successful baryogenesis.
In the two Higgs doublet model,
the magnitude of the deviation from the standard model value 
is shown to be larger than 10\%.
Such a deviation can be detected at a future electron-positron linear collider.
\end{abstract}
%
\maketitle
\thispagestyle{fancy}
%
%
%
\section{INTRODUCTION} 
The Higgs boson is responsible for
the spontaneous breakdown of the electroweak symmetry, which leads to
masses for the gauge bosons and quarks/leptons.
The direct search of the Higgs boson at the LEP and various electroweak observable data
indicate that the mass of the standard model (SM) Higgs boson is
larger than 114 GeV~\cite{Barate:2003sz} and less than 285 GeV~\cite{lep-ewwg}.
Such a light Higgs boson can be discovered at the CERN LHC.
The profile of the Higgs boson is expected to be throughly determined 
at the international Linear Collider (ILC) as well as the LHC.
At the ILC, the Higgs couplings to the gauge bosons and the heavy fermions
($b$ quark, $\tau$ lepton) 
can be measured with high accuracy, and determination of the triple Higgs boson coupling 
is expected to be ${\cal O}(10-20)\%$ level~\cite{hhh-exp}.
The high energy collider experiments will not only reveal the particle physics at the TeV scale,
but also open the window to cosmological connection to particle physics, 
such as dark matter and bayogenesis.
\\
\indent
In this talk, we focus on a collider signature of electroweak baryogenesis~\cite{kos}.
It is observed that the ratio of 
the baryon number density to entropy density is $n_B/s\sim 10^{-10}~$\cite{wmap}.
According to the Sakharov's criteria~\cite{Sakharov:1967dj}, there are three requirements for
generation of the baryon asymmetry: (a) baryon number violation, (b) $C$ and $CP$ violation,
and (c) deviation from thermal equilibrium.
It is well known that the condition (c) is not satisfied in the SM with 
the current Higgs mass bound, 
and that the $CP$ violating phase in the Kobayashi-Maskawa matrix
is too small to generate the sufficient baryon asymmetry. 
Here, we consider electroweak baryogenesis 
in the two Higgs doublet model (2HDM)~\cite{ewbg-thdm, simplified-thdm, cline, funakubo}
and the minimal supersymmetric standard model (MSSM)~\cite{ewbg-mssm}.
In particular, we study a relationship between the strength of
the electroweak phase transition and 
the quantum corrections to the trilinear coupling of the lightest Higgs boson.
%
%
\section{ELECTROWEAK PHASE TRANSITION IN THE 2HDM}
The 2HDM is a simple extension of the SM by adding the second Higgs doublet.
In this model, the $Z_2$ symmetry ($\Phi_1\rightarrow\Phi_1,~
\Phi_2\rightarrow-\Phi_2$) is imposed on the Yukawa interactions
to avoid the tree-level Higgs-mediated flavor changing neutral current processes~\cite{fcnc}.
Consequently, the Higgs potential at the tree-level takes the form 
\begin{eqnarray}
V(\Phi_1, \Phi_2)&=&m_1^2|\Phi_1|^2+m_2^2|\Phi_2|^2
	-(m_3^2\Phi_1^\dagger\Phi_2+\mbox{h.c.})\non\\
	&&+\frac{\lambda_1}{2}|\Phi_1|^4+\frac{\lambda_2}{2}|\Phi_2|^4
	+\lambda_3|\Phi_1|^2|\Phi_2|^2+\lambda_4|\Phi_1^\dagger\Phi_2|^2
	+\frac{1}{2}\Big[\lambda_5(\Phi_1^\dagger\Phi_2)^2+\mbox{h.c.}\Big],
\end{eqnarray}
where $m_3^2$ or $\lambda_5$ can be complex.
Here, we assume that their phases are small and neglect them at the first approximation. 
To simplify our analysis we consider the phase transition in the direction of
$\langle\Phi_1\rangle=\langle\Phi_2\rangle=(0~\varphi)^T/2$,
which corresponds to $m_1=m_2,~\lambda_1=\lambda_2$, in other words,
$\sin(\beta-\alpha)=\tan\beta=1$~\cite{simplified-thdm, cline}.\\
\indent
The one-loop contributions to the effective potentials at zero and finite temperatures
~\cite{dj} are respectively given by
\begin{equation}
V_1(\varphi)=n_i\frac{m_i^4(\varphi)}{64\pi^2}
	\bigg(\log\frac{m_i^2(\varphi)}{Q^2}-\frac{3}{2}\bigg),\quad
V_1(\varphi, T)=\frac{T^4}{2\pi^2}
	\Big[\sum_{i={\rm bosons}}n_iI_B(a^2)+n_tI_F(a^2)\Big],\label{eff-pot}
\end{equation}
with
\begin{equation}
I_{B, F}(a^2)=\int_0^\infty dx~x^2\log\Big(1\mp e^{-\sqrt{x^2+a^2}}\Big),
\quad a(\varphi)=\frac{m(\varphi)}{T},
\end{equation}
where $Q$ is a renormalization scale, $m_i(\varphi)$ is the field dependent mass of
the particle $i$,
and $n_i$ is the degrees of the freedom of $i$, i.e.,
$n_W=6,~n_Z=3$ for gauge bosons ($W^\pm$, $Z$), $n_t=-12$ for top quark($t$)
and $n_h=n_H=n_A=1,~n_{H^\pm}=2$ for the five physical Higgs bosons ($h, H, A, H^\pm$). 
\\
\indent
The qualitative features of the phase transition can be understood by the following 
high temperature expansion. 
When $m_{\Phi}^2\gg m_h^2,~M^2~(\Phi\equiv H, A, H^\pm,~M^2
\equiv m_3^2/\sin\beta\cos\beta)$, the field dependent masses
of the heavy Higgs bosons can be written as $m_\Phi^2(\varphi)\simeq m_\Phi^2\varphi^2/v^2$.
At high temperatures, the Higgs potential can be expanded in powers of $\varphi$~\cite{analy-exp}.
\begin{equation}
V_{\rm eff}\simeq D(T^2-T_0^2)\varphi-ET\varphi^3+\frac{\lambda_{T}}{4}\varphi^4
	+\cdots,\label{Veff}
\end{equation}
where
$E=\frac{1}{12\pi v^3}(6m_W^3+3m_Z^2+m_H^3+m_A^3+2m_{H^\pm}^3)$.\label{cube}
The non-zero $E$ makes the phase transition first order. 
In order to preserve the generated baryon asymmetry, the sphaleron process
must decouple after the phase transition. This condition gives~\cite{sph}
\begin{equation}
\frac{\varphi_c}{T_c}\gtsim1,\label{sph}
\end{equation}
where $\varphi_c$ is the vacuum expectation value of the Higgs boson at
the critical temperature $T_c$.
One can easily see
\begin{equation}
\varphi_c=\frac{2ET_c}{\lambda_{T_c}},
\end{equation}
where $\lambda_{T_c}$ is the quartic coupling  at $T_c$.
Due to the contributions of the heavy Higgs bosons in the loop, 
the first order phase transition can be strong enough to satisfy Eq.~(\ref{sph}).
The high temperature expansion makes it easy to see the 
phase transition analytically.
However, it breaks down when the
masses of the particles in loops become larger than $T_c$.
In the following, we therefore calculate $T_c$ and $\varphi_c$ numerically.
%
%
\section{ RADIATIVE CORRECTIONS TO THE TRILINEAR COUPLING}
We also calculate the trilinear coupling of the lightest Higgs boson (the $hhh$ coupling)
at zero temperature in the parameter region where the phase transition is strongly first order.
The leading contribution of the heavy Higgs bosons and the top quark
to the $hhh$ coupling
can be extracted from the one-loop calculation by \cite{hhh-coupling}
\begin{eqnarray}
\lefteqn{\lambda_{hhh}^{\rm eff}(\mbox{2HDM})}\non\\
&\simeq&\frac{3m_h^2}{v}
	\Bigg[1+\frac{m_H^4}{12\pi^2m_h^2v^2}\Bigg(1-	\frac{M^2}{m_H^2}\Bigg)^3
	+\frac{m_A^4}{12\pi^2m_h^2v^2}\Bigg(1-\frac{M^2}{m_A^2}\Bigg)^3
	+\frac{m_{H^\pm}^4}{6\pi^2m_h^2v^2}
	\Bigg(1-\frac{M^2}{m_{H^\pm}^2}\Bigg)^3-\frac{m_t^4}{\pi^2m_h^2v^2}\Bigg].
	\label{lambdahhh}
\end{eqnarray}
It is easily seen that the effects of the heavy Higgs boson loops are enhanced 
by $m_\Phi^4$ ($\Phi=H, A, H^\pm$) when $M^2$ is zero.
These effects do not decouple in the 
large mass limit $m_\Phi \rightarrow \infty$ and yields the large deviation of the
$hhh$ coupling from the SM prediction.
In this case, $m_\Phi$ is bounded from above by perturbative unitarity
($m_\Phi\ltsim 550$ GeV)~\cite{pub-2hdm}.
We note that when such nondecoupling loop effects due to the extra heavy 
Higgs bosons are large on the $hhh$ coupling, the coefficient $E$ of the cubic term 
in Eq~(\ref{Veff}) becomes large in this model. 
Therefore there is a strong correlation between the large quantum correction
to the $hhh$ coupling and successful electroweak baryogenesis.
%
%
\section{NUMERICAL EVALUATION}
\begin{figure}[t]
\centerline{\includegraphics[width=15cm]{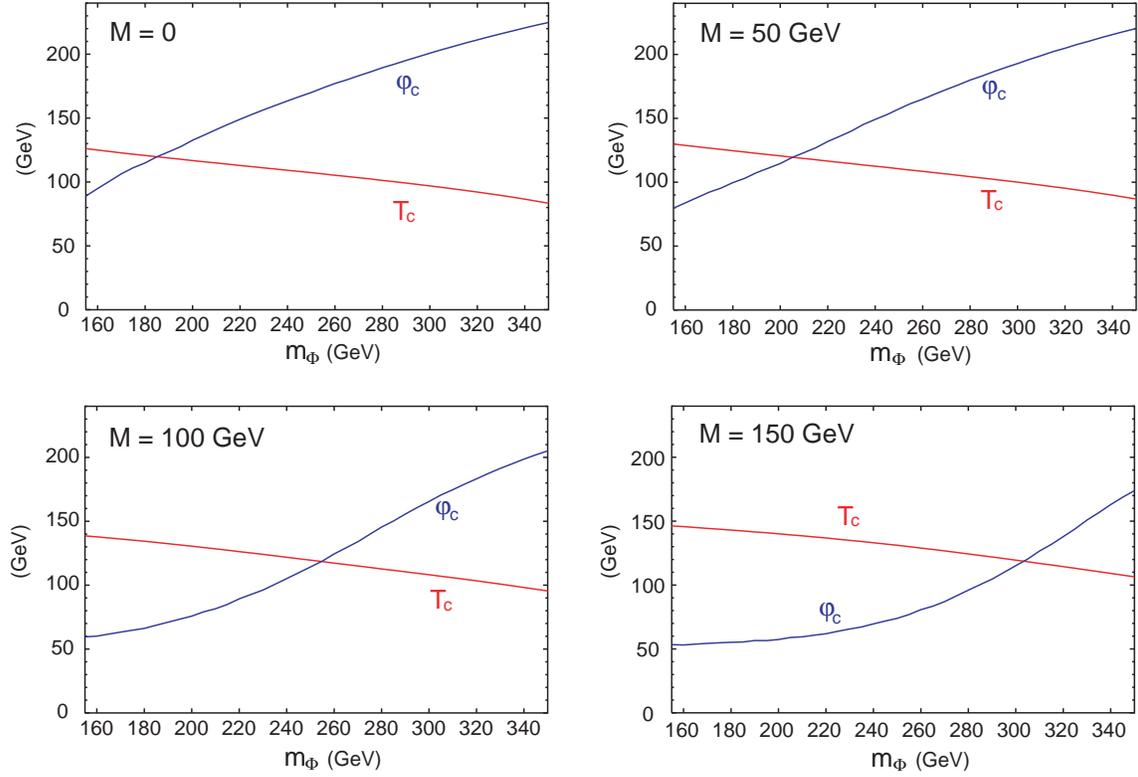}}
\caption{The Higgs vacuum expectation value $\varphi_c$ at the critical temperature $T_c$
as a function of the heavy Higgs boson mass $m_\Phi^{}$ ($m_\Phi=m_H=m_A=m_{H^\pm}$)
for $M=0$, 50, 100 and 150 GeV. Other parameters are fixed as
$\sin(\alpha-\beta)=\tan\beta=1$ and $m_h=120~\mbox{GeV}$.} \label{fig1-phi_c-T_c}
\end{figure}

We calculate the effective 
potential (\ref{eff-pot}) varying the temperature $T$
and determine the critical temperature $T_c$ of the first order
phase transition and the expectation value $\varphi_c$ at $T_c$. 
Figs. \ref{fig1-phi_c-T_c} show
the $T_c$ and $\varphi_c$
as a function of the mass of the heavy Higgs boson $m_\Phi$ 
for $M=0,~50,~100$ and 150 GeV.
We take $\sin(\alpha-\beta)=-1$, $\tan\beta=1$ and $m_h=120$ GeV.
For the heavy Higgs boson mass, we assume
$m_H^{}=m_{A}^{}=m_{H^\pm}^{} (\equiv m_\Phi^{})$ to avoid the constraint
on the $\rho$ parameter from the LEP precision data~\cite{pdg}.
We also take into account the ring summation for
the contribution of the Higgs bosons to the effective potential at finite temperature
to improve our calculation~\cite{dj,daisy}.
In the case of $M=0$, it is found that $\varphi_c=T_c\simeq120$ GeV
at $m_\Phi\simeq185$ GeV, and the condition (\ref{sph}) is 
satisfied for $m_\Phi\gtsim 185$ GeV.
One can also find that the condition (\ref{sph}) can still be satisfied for $M=150$ GeV, 
if the masses of the heavy Higgs bosons are greater than about 300 GeV.
%
%
\\
\indent
\begin{figure}[t]
\centerline{\includegraphics[width=16cm]{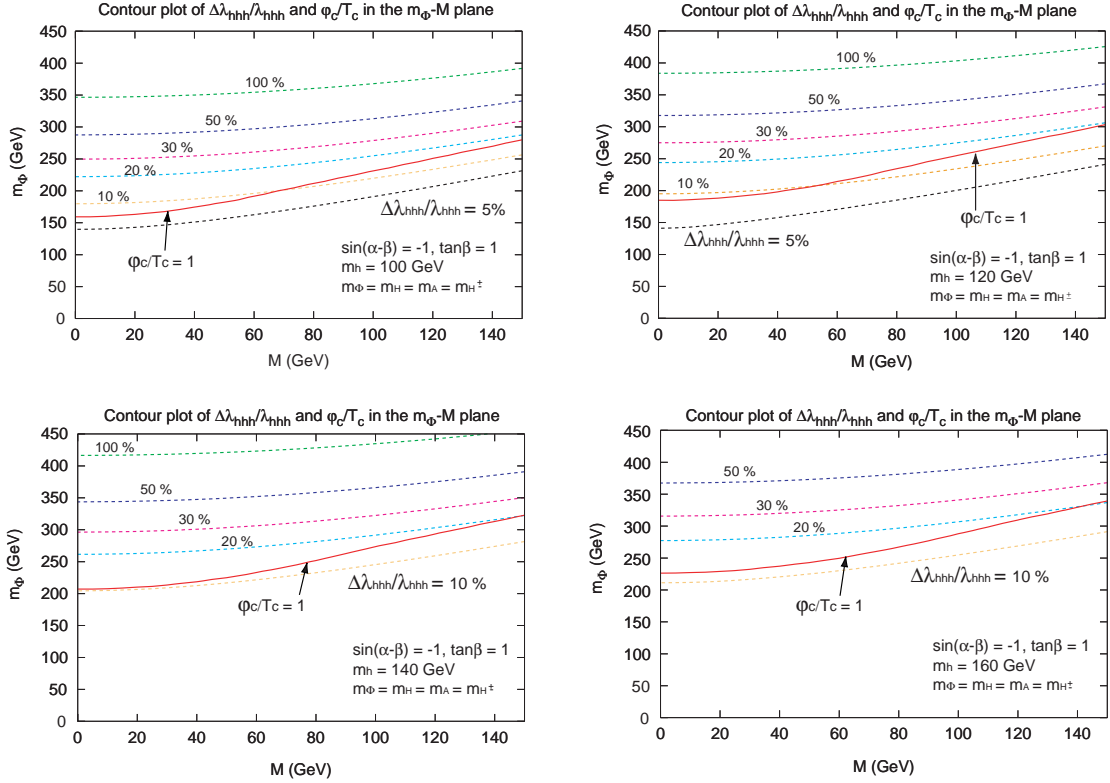}}
\caption{The contour of the radiative correction of the triple Higgs boson coupling
constant overlaid with the line $\varphi_c/T_c=1$ in the $m_\Phi^{}$-$M$ plane
for $m_h$=100, 120, 140 and 160 GeV.
Other parameters are the same as those in Fig.~\ref{fig1-phi_c-T_c}.
The above the critical line, the phase transition is strong enough for the
successful electroweak baryogenesis scenario.}\label{fig1}
\end{figure}
In Figs.~\ref{fig1},
we show the parameter region where the necessary condition of electroweak
baryogenesis in Eq.~(\ref{sph}) is satisfied in the $m_\Phi^{}$-$M$ plane
for $m_h=100,~120,~140$ and 160 GeV.
We also take $\sin(\alpha-\beta)=-1$ and $\tan\beta=1$.
For $m_h=120$ GeV, we can see that the phase transition becomes strong
enough for successful baryogenesis when the masses
of the heavy Higgs bosons are larger than about 200 GeV. 
For the larger values of $M$ or $m_h$, the greater
$m_\Phi^{}$ are required to satisfy the condition (\ref{sph}). 
In this figure we also plot the contour of the
magnitude of the deviation in the $hhh$ coupling from the SM value.
We define the deviation 
$\Delta \lambda_{hhh}^{\rm 2HDM}/
\lambda_{hhh}^{\rm eff}(\rm SM)$ 
by $\Delta \lambda_{hhh}^{\rm 2HDM}\equiv
\lambda_{hhh}^{\rm eff}(\rm 2HDM)-\lambda_{hhh}^{\rm eff}(\rm SM)$.
We calculated the deviation at the one loop level in the on-shell scheme
which gives a better approximation than the formula given in Eq. (\ref{lambdahhh})
~\cite{hhh-coupling}.
We can easily see that the magnitude of the deviation is significant ($\gtsim10\%$)
in the parameter region where the electroweak baryogenesis is possible. 
Such magnitude of the deviation can be detected at a future LC experiment.
\\
\indent
Next we discuss a scenario of electroweak baryogenesis in the MSSM.
The strong first order phase transition can be induced by the loop effect
of the light stop in the finite temperature effective potential~\cite{ewbg-mssm}.
We examine the loop effect of the light stop on the $hhh$ coupling in this scenario.
In the following, we only consider the finite and zero temperature effective potentials
using high temperature expansion to understand the qualitative feature.
As we have done in the case of the 2HDM, 
we consider the relationship between the magnitude of the 
phase transition and the deviation of the $hhh$ coupling from the SM value. 
The combined result is approximately expressed as
\begin{equation}
\frac{\Delta \lambda_{hhh}({\rm MSSM})}{\lambda_{hhh}({\rm SM})} 
 \simeq \frac{2v^4}{m_t^2m_h^2}(\Delta E_{\tilde{t}_1})^2,
\end{equation}
where $m_h$ is the one-loop renormalized mass of the lightest Higgs boson and 
$\Delta E_{\tilde{t}_1}$ is the contribution of the light stop loop to the cubic term
in the finite temperature effective potential.
From the condition (\ref{sph}), the deviation in the $hhh$ coupling
from the SM value is estimated to be $\sim6\%$ for $m_h=120$ GeV. 
In the MSSM, the condition of the sphaleron decoupling also leads to the large deviation
of the $hhh$ coupling from the SM prediction at zero temperature.
\\
\indent
In this talk, we have investigated that a phenomenological
consequence of electroweak baryogenesis in the 2HDM and the MSSM.
We found that the evidence of electroweak baryogenesis
appears in the $hhh$ coupling constant in both models. 
In the 2HDM, the magnitude of such a coupling constant
can deviate from the SM prediction enough to be detected at the ILC.
\begin{acknowledgments}
The work of YO was supported in part by a Grant-in-Aid of the Ministry 
of Education, Culture, Sports, Science, and Technology, Government of
Japan, Nos.~13640309, 13135225, and 16081211.
\end{acknowledgments}
%
%


%
\end{document}